**On the scope of applicability of the models of Darwinian dynamics**
G. Karev

National Center for Biotechnology Information, National Institutes of Health,
Bldg. 38A, 8600, Rockville Pike, Bethesda, MD 20894, USA; gkarev@gmail.com

**Abstract**
In their well-known textbook (Vincent & Brown, 2005), Vincent and Brown suggested an attractive approach for studying evolutionary dynamics of populations that are heterogeneous with respect to some strategy that affects the fitness of individuals in the population. The authors developed a theory, whose goal was to expand the applicability of mathematical models of population dynamics by including dynamics of an evolving heritable phenotypic trait subject to natural selection. The authors studied both the case of evolution of individual traits and of mean traits in the population (or species) and the dynamics of total population size.
The authors consider the developed approach as (more or less) universally applicable to models with any fitness function and any initial distribution of strategies, which is symmetric and has small variance.
Here it was shown that the scope of the approach proposed by Vincent & Brown is unfortunately much more limited. I show that the approach gives exact results only if the population dynamics linearly depends on the trait; examples where the approach is incorrect are given.

**1. Introduction.**
In their well-known book, Vincent and Brown (Vincent & Brown, 2005) developed a theory, whose goal was to extend applicability of mathematical models of population dynamics by including the dynamics of an evolving heritable phenotypic trait subject to natural selection. The authors considered both the case of evolution of individual traits and of mean traits in the population (or species) and the dynamics of total population size. Similar problems were studied in replicator dynamics (Hofbauer & Sigmund, 1998), (Kareva & Karev, 2019), with the results being similar in some simple cases and different in other cases. Below I try to reveal the reasons of these differences.

Vincent and Brown's theory was formulated in terms of evolutionary game theory that was initially developed by Maynard Smith and Price (Smith & Price, 1973; Smith, 1982). The models considered by the authors are composed of two parts: population dynamics that are governed by fitness, which in turn is influenced by changes in mean strategy, and the mean strategy dynamics that can change over time as influenced by population dynamics and changes in population composition. Both population density and mean strategy influence fitness of individuals.

Let us describe the approach and the model of Vincent and Brown (hereafter termed the V&B model) in a simple case. Consider a population composed of individuals with *n* different strategies, such that each strategy can affect the individual's fitness. Let $\mathbf{U} = \{u_i, i = 1, \dots n\}$ be the set of (scalar) strategies $u_i$, which represent a heritable characteristic of an individual. Let $l_i$ be the density of individuals who possess strategy $u_i$ and compose, by definition, the *i*-th species (or phenotype). All individuals in the population are represented by vector $\mathbf{L} = [l_1, \dots l_n]$, and all the strategies currently in the population are given by vector $\mathbf{u} = [u_1, \dots u_n]$.

Let $H_i(\mathbf{u}, \mathbf{L})$ be the individual fitness for the species *i* given densities and strategies of all species in the population, so that population dynamics are given by equation

$$\frac{dl_i}{dt} = l_i H_i(\mathbf{u}, \mathbf{L}). \tag{1.1}$$

The well-known equations for frequencies and total population size are given as equations (4.9) - (4.12) in V&B.

A central new notion introduced by V&B and used throughout the textbook is that of a "fitness generating function". By definition, a function $G(v, \mathbf{u}, \mathbf{l})$ is a fitness generating function (G-function) for the population dynamics if and only if

$$G(v, \mathbf{u}, \mathbf{L})|_{v=u_i} = H_i(\mathbf{u}, \mathbf{L}). \tag{1.2}$$

Similar definitions for more complex situations were also given and used both in V&B textbook, as well as in subsequent publications by other teams. For example, in (R. Rael, Costantino, Cushing, & Vincent, 2009) the authors studied, based on the V&B theory, a model in which strategies are not phenotypes but allele frequencies; in (Pressley et al., 2021) and (Martınez et al., 2022), the authors studied a cancer model where strategies are defined as resistance to therapy.

Everywhere, where the term "strategy" refers to individual strategy (phenotype, trait), the fitness generating function is just a convenient way of expressing all individual fitness functions in one expression. It becomes non-trivial when the notion of G-function is extended by using the mean value of strategies within a species or population instead of an individual's strategy. V&B proposed equations that describe dynamics of a population and the mean strategy within it and compose their model of Darwinian dynamics in different permutations; let us emphasize that these equations use the G-function on mean value of strategies.

The authors start from the model of a finite number of species and a finite number of possible strategies for all individuals within each species (see V&B, equations (5.6) and (5.9)). Regrettably, the derivation of these equations was not entirely correct even in this simple case.

The use of the mean strategy in the G-function instead of individual strategies is a source of possible errors; examples are given below.

**2. The model of "Darwinian dynamics" does not follow from phenotype dynamics**
Let us consider these calculations in detail.
According to V&B, page 121:
> "Strategy dynamics is derivable from the population dynamics equations provided that a distribution of strategies about some mean exists for each species. This distribution requires a distinction between phenotypes and species…. We refer to a **species** as a set of evolutionarily identical individuals whose strategies, referred to as **phenotypes**, aggregate around a distinct mean strategy value… We show below that, when the phenotypes of a particular species aggregate around some mean strategy value, the distribution of strategies results in an evolutionary dynamic. "
>
> "In order to simplify what follows we use the following short-cut notation
> $G|_w = G(v, \mathbf{u}, \mathbf{L})|_{v=w}$,
> $\frac{\partial G}{\partial v}|_w = \frac{\partial G(v,\mathbf{u},\mathbf{L})}{\partial v}|_{v=w}$ (2,1)

where *w* is any scalar strategy. Using this notation, the population dynamics for this category of *G*-functions becomes the following

$$dx_i/dt = x_i \, G|_{u_i}. \tag{2.2}$$

"The variable $u_i$ no longer refers to the strategy of any given individual in the population $x_i$, but rather we now define $u_i$ to be the **mean strategy** of all individuals in the population $x_i$."

Let us emphasize that here the population $x_i$ is inhomogeneous and populations differ from each other only by the mean strategy; neither variance nor other characteristics of strategy distribution within populations are taken into account.

Equation (2.2) is presented by the authors as evident corollary of the original equation (1.1) and notation (2.1). Below I show that (2.2) **does not follow** from (1.1) and (2.1) when $x_i$ refers to the density of inhomogeneous *i*-th population composed of different species and $u_i$ refers to not an individual strategy, but to the mean value of strategies (or phenotypes) within the *i*-th population.

V&B studied at first the simple case of a population composed of a finite number of species and a finite number of possible phenotypes. They wrote:

"We use the notation $x_{ij}$ to designate a phenotype *j* within the species *i*. It follows from the definition of G-function that the population dynamics for the phenotypes are given by the following:

$$\frac{dx_{ij}}{dt} = x_{ij} \, G|_{u_{ij}}. \tag{2.3}$$

The density of *i*-th species $x_i = \sum_j x_{ij}$; $u_{ij}$ designates the strategy used by the phenotype $x_{ij}$, and $u_i = (\sum_j x_{ij} u_{ij})/x_i$ is the mean value of strategies within the *i*-th species."

Equation (2.3), which describes "phenotype dynamics" (or dynamics of *homogeneous species* $x_{ij}$ composed of identical individuals using strategy $u_{ij}$) as the starting point, the dynamics of densities of *inhomogeneous* species $x_i$ and the *mean values* of phenotypes (strategies) $u_i$ should be derived from this initial system.

In order to make the problem as transparent as possible, I will consider the simplest case of only a single inhomogeneous species (so the index $i$ takes a single value and may be omitted); the species is composed of individuals that can have $n$ different phenotypes.

It seems that using identical notation for different quantities, such as both for individual strategy and for the **mean strategy** of individuals within the species, may be the source of confusion. To avoid this issue, I will use different notation for different quantities.

Let $\mathbf{L} = [l_1, \ldots l_n]$ be the vector of current densities of phenotypes and let $\mathbf{v} = [v_1, \ldots v_n]$ be the vector of corresponding strategies (i.e., phenotype $l_j$ uses strategy $v_j$). Notice that the set of possible strategies or phenotypes is fixed for every model, so $\mathbf{v}$ is not a variable (and in what follows it may be omitted).

Let us denote $l_j(t)$ to describe the density of $j$–th phenotype using strategy $v_j$ at time $t$. The total population of the species is $\sum_j l_j$, the frequency of phenotypes $l_j$, or the frequency of the strategy $v_j$ is $p_j = l_j/x$; the mean strategy in the species at $t$ moment is

$$u(t) \equiv E^t[v] = \sum_j p_j(t) v_j .$$

According to (2.3) and (1.1), the dynamics for the phenotypes are given by the following equations:

$$\frac{dl_j}{dt} = l_j H_j(\mathbf{L}) \equiv l_j \, G|_{v_j}. \tag{2.4}$$

Here the fitness-generating function $G(v, \mathbf{L})$ is defined by equation (1.2) that now has the form

$$G(v, \mathbf{L})|_{v=v_j} = H_j(\mathbf{L}).$$

Equation (2.4) describes the initial dynamics of all the phenotypes, and all the equations for the dynamics of the total system, including equations for total population size, mean and variance of strategy distribution, etc., should be derived from it.

The equations of the Darwinian dynamics in V&B (5.1), (5.6), (5.9) in case of a single species read (up to notation)

$$\frac{dx}{dt} = x \, G(u, \mathbf{L}) \tag{2.5a}$$

$$\frac{du}{dt} = \sigma^2(t) \frac{\partial G}{\partial v}(v, \mathbf{L})|_{v=u} \tag{2.5b}$$

where $u(t) = E^t[v]$ and $\sigma^2(t)$ is the variance of the strategy's distribution.

The following intermediate equation was used by V&B to derive the last equation:

$$\frac{du}{dt} \equiv \frac{d}{dt} E^t[v] = \sum_j v_j p_j [G(v_j, \mathbf{L}) - G(u, \mathbf{L})] \tag{2.6}$$

Below I show that equations (2.5a), (2.5b), (2.6) in general case cannot be derived from and may even contradict the initial dynamical system (2.4) and therefore in general can be wrong; similarly, equations (5.1), (5.6), (5.9) in V&B cannot be derived from the initial model V&B (5.2).

Let me start from the equation for population density (2.5a). Using the initial equation (2.4) for dynamics of the phenotypes results in the following:

$$\frac{dx}{dt} = \frac{d \sum_j l_j}{dt} = \sum_j l_j(t) H_j(\mathbf{L}) = x(t) \sum_j p_j(t) G(v_j, \mathbf{L}) = x(t) E^t[G(v, \mathbf{L})] \tag{2.7}$$

where $x(t)$ is the total population size and

$$E^t[G(v, \mathbf{L})] = \sum_j p_j(t) G(v_j, \mathbf{L}) = \sum_j p_j H_j(\mathbf{L})$$

is the mean value of the phenotype's fitness. (Notably, this equation in different notation was given on p.105 of V&B as Equation (4.12)).

In contrast, equation (2.5a) states that

$$\frac{dx(t)}{dt} = x(t) \, G(E^t v, \mathbf{L}) = x(t) G(\sum_j p_j(t) v_j, \mathbf{L}).$$

However,

$$E^t[G(v, \mathbf{L})] = \sum_j p_j \, G(v_j, \mathbf{L}) \neq G(\sum_j p_j v_j, \mathbf{L}) = G(E^t[v], \mathbf{L}), \tag{2.8}$$

unless the distribution of $v$ is concentrated in a single point or, in case of a general distribution, if the function $G$ is linear over the first argument, i.e. if

$$G(v, \mathbf{L}) = v G_1(\mathbf{L}) + G_2(\mathbf{L}). \tag{2.9}$$

Inequality in (2.8) just means the well-known fact that the mean value of a function on a random variable in general **is not equal** to that function on the mean value of that random variable, i.e., in our case that

$$E^t[G(v, \mathbf{L})] \neq G(E^t[v], \mathbf{L}) \tag{2.10}$$

The equality in (2.10) for non-degenerate (i.e. not concentrated in a single point) distributions holds *only for linear* functions (2.9).

So, in a general case, Equation (2.5a) suggested by V&B for dynamics of population size (see V&B Eq 5.1) does not follow from and contradicts initial equations (1.1) and (2.3), which describe the dynamics of species that compose the population even in the case of a single population (2.4).

It is a grave problem in the derivation of equations of Darwinian dynamics, which is implicitly repeated in the derivation of the corresponding equations for different Darwinian models of population dynamics in subsequent chapters of V&B.

**Example 1.** Let $\frac{dl_j(v)}{dt} = l_j v^2$, so that $G(v, \mathbf{L}) = v^2$, and let $v_j = j$, $p_j = \frac{1}{2}, j = 1,2$. Then $E^t[G(v, \mathbf{L})] = \sum_j p_j\, G(p_j, \mathbf{L}) = \frac{1}{2}\sum_{j=1}^{2} j^2 = 5/2$ ; however, $G(E^t[v], \mathbf{L}) = \left(\frac{1}{2}\sum_{j=1}^{2} j\right)^2 = 9/4$.

The correct equation for dynamics of total species density is well-known in replicator dynamics; in terms of model (2.4), it is Equation (2.7). Let us emphasise again that (the correct) equation (2.7) is not equivalent to equation (2.5a) because
$$E^t[G(v, \mathbf{L})] \neq G(E^t[v], \mathbf{L})$$
unless the function $G$ is linear with respect to $v$.

Now let us estimate the difference between V&B equation (2.5a) and the correct equation (2.7), i.e. the difference $E^t[G(v, \mathbf{l})] - G(E^t[v], \mathbf{l})$.

Expanding the function $G(v, \mathbf{l})$ in Taylor series over $v$ at the point $u = E^t[v]$, we get
$$G(v, \mathbf{L}) = G(u, \mathbf{L}) + (v - u)\frac{\partial}{\partial v} G(v, \mathbf{L})|_{v=u} + \frac{(v-u)^2}{2}\frac{\partial^2}{\partial v^2} G(v, \mathbf{L})|_{v=u} + O((v-u)^3).$$
Following V&B (p.121) let us assume that "the phenotypes of a particular species aggregate around some mean strategy value"; then we get
$$E^t[G(v, \mathbf{L})] - G(E^t[v], \mathbf{L}) \cong E^t\left[\frac{(v-u)^2}{2}\frac{\partial^2}{\partial v^2} G(v, \mathbf{L})|_{v=u}\right] = \frac{\sigma^2(t)}{2}\frac{\partial^2}{\partial v^2} G(v, \mathbf{L})|_{v=u}.$$
Consequently, instead of (incorrect) equation (2.5a), one should use either Equation (2.7), or the approximate equation of the 2nd order
$$\frac{dx}{dt} \cong x\left(G(u, \mathbf{L}) + \frac{\sigma^2(t)}{2}\frac{\partial^2}{\partial v^2} G(v, \mathbf{L})|_{v=u}\right). \tag{2.11}$$
Notice that the last equation implies the V&B equation (2.5a) only if either $\sigma^2(t) \cong 0$, i.e. the distribution of strategies is concentrated in (or around) a single point *at any time*, or if $\frac{\partial^2}{\partial v^2} G(v, \mathbf{L})|_{v=u} \cong 0$, i.e. the function $G(v, \mathbf{L})|_{v=u}$ is (almost) linear.

**Example 2.** Let $\frac{dl_j(v)}{dt} = l_j v^2$, so $G(v, \mathbf{v}, \mathbf{L}) = v^2$. Then equation (2.5a) becomes

$$\frac{dx}{dt} = xG(u, \mathbf{l}) = xu^2,$$

while equation (2.11) becomes

$$\frac{dx}{dt} \cong x\left(G(u, \mathbf{L}) + \frac{\sigma^2(t)}{2}\frac{\partial^2}{\partial v^2}G(v, \mathbf{L})|_{v=u}\right) = x(u^2 + \sigma^2). \tag{2.12}$$

The correct equation (2.7) is

$$\frac{dx(t)}{dt} = x(t)E^t[G(v, \mathbf{L})] = x(t)E^t[v^2] = x(t)E^t[v^2 - u^2 + u^2] = x(u^2 + \sigma^2).$$

Therefore, the approximate equation (2.12) coincides in this case with the correct equation.

**Example 3a.** Let $\frac{dl_j(v)}{dt} = l_j v^k$, so $G(v, \mathbf{v}, \mathbf{L}) = v^k, k > 1$. Then equation (2.5a) reads

$$\frac{dx}{dt} = x\, G(u, \mathbf{L}) = xu^k,$$

while equation (2.12) becomes

$$\frac{dx}{dt} \cong x\left(G(u, \mathbf{l}) + \frac{\sigma^2(t)}{2}\frac{\partial^2}{\partial v^2}G(v, \mathbf{L})|_{v=u}\right) = x(u^k + \frac{\sigma^2}{2}k(k-1)u^{k-2})$$

$$\text{(2.12.b)}$$

The correct equation (2.7) is

$$\frac{dx(t)}{dt} = x(t)E^t[G(v, \mathbf{L})] = x(t)E^t[v^k].$$

That is, if one expands the right-hand side of the last equation to the Taylor series around $E^t[v] = u$ up to the term of the 2nd order, they will get the approximate equation (2.12b).

**Example 3b.** Now let $\frac{dl_j(v)}{dt} = l_j \exp(-av)$, so $G(v, \mathbf{L}) = \exp(-av)$.

Then equation (2.5a) becomes

$$\frac{dx}{dt} = x\, G(u, \mathbf{L}) = x\exp(-au),$$

while equation (2.12) becomes

$$\frac{dx}{dt} \cong x\left(G(u, \mathbf{L}) + \frac{\sigma^2(t)}{2}\frac{\partial^2}{\partial v^2}G(v, \mathbf{L})|_{v=u}\right) = x\exp(-au)(1 + \frac{\sigma^2(t)}{2}a^2).$$

The correct equation (2.7) in this case is

$$\frac{dx(t)}{dt} = x(t)E^t[G(v, \mathbf{L})] = x(t)E^t[\exp(-av)] = x(t)L_t(a),$$

where $L_t$ is the Laplace transform of the distribution of strategies *v* at *t* moment, which may be different depending on initial distribution.

So, in this example, both equations, (2.5a) and (2.12), are incorrect.

Next, I will consider the derivation of the second equation (2.5b) of Darwinian dynamics for the mean strategy. At first, let me show that the intermediate equation (2.6) that was used by V&B for derivation of equation (2.5b) is not correct; moreover, it contains an internal contradiction. Indeed, the last equation on page 123 can be rewritten in our notation as

$$\frac{dp_j}{dt} = \frac{d}{dt}\frac{l_j}{x} = \frac{xl_j' - l_j x'}{x^2} = \frac{xl_j G(v_j, \mathbf{l}) - l_j xG|_u}{x^2} = p_j(G(v_j, \mathbf{L}) - G(u, \mathbf{L})). \quad (2.13)$$

Notice that two equations were used here:

$$l_j' = l_j G(v_j, \mathbf{L}) \text{ and } x' = xG|_u.$$

However, these equations contradict each other; indeed, if the 1st equation holds, then

$$x' = \sum_j l_j' = \sum_j l_j G(v_j, \mathbf{L}) = x\sum_j p_j G(v_j, \mathbf{L}) = x E^t[G(v, \mathbf{L})] \neq xG(E^t[v], \mathbf{L}) = xG|_u$$

unless the function *G* is linear over *v*.

Let us emphasize that equation (2.13) for frequencies *cannot be valid* for all *j* unless

$$E^t[G(v, \mathbf{L})] = G(E^t[v], \mathbf{L}).$$

Indeed, as $\sum_j p_j = 1$, it should be $\sum_j \frac{dp_j}{dt} = 0$. According to equations (2.13) and (2.11)

$$\sum_j \frac{dp_j}{dt} = E^t[G(v, \mathbf{l})] - G(E^t[v], \mathbf{l}) \cong \frac{\sigma^2(t)}{2}\frac{\partial^2}{\partial v^2}G(v, \mathbf{l})|_{v=u}$$

The left-hand of the last equation should be equal to 0, but its right-hand side is equal to 0 only if $\sigma^2(t)=0$, that is, if there exists only a single value of the strategy, or if the function $G(v, \mathbf{l})$ is linear over *v*.

The correct equation for frequencies is

$$\frac{dp_j}{dt} = \frac{d}{dt}\frac{l_j}{x} = \frac{xl_j' - l_j x'}{x^2} = \frac{l_j G(v_j, \mathbf{L})}{x} - p_j\frac{x'}{x} = p_j(G(v_j, \mathbf{L}) - E^t[G(v, \mathbf{L})]) \quad (2.14)$$

(see also V&B, equations (4.9)-(4.10)).

So, due to inequality (2.10),

$$\frac{dp_j}{dt} \neq p_j\left(G(v_j, \mathbf{L}) - G(E^t[v], \mathbf{L})\right),$$

and the equality holds only if the function $G(v, \mathbf{L})$ is linear over $v$; in the general case, the equality is incorrect (a similar equation in the case of multiple species was given by V&B, page 123).

Next, by definition, the mean strategy is

$u(t) = E^t[v] = \sum_j p_j(t) v_j$, so

$$\frac{dE^t[v]}{dt} = \sum_j v_j \frac{dp_j}{dt} = \sum_j p_j v_j (G(v_j, \mathbf{L}) - E^t[G(v, \mathbf{L})]) \tag{2.15}$$

Hence, equation (2.6) does not hold because in the general case

$$\sum_j p_j v_j G(v_j, \mathbf{v}, \mathbf{L}) - E^t[G(.,\mathbf{v}, \mathbf{L})]) \neq \sum_j p_j v_j \left( G(v_j, \mathbf{v}, \mathbf{L}) - G(E^t[v], \mathbf{v}, \mathbf{L}) \right).$$

So, the *correct* first-order approximation

$$G(v_j, \mathbf{L}) - G(E^t[v], \mathbf{L}) \approx \frac{\partial G(E^t[v], \mathbf{L})}{\partial v} \delta v_j \text{ where } \delta v_j = v_j - E^t[v],$$

applied to the *incorrect* equation

$$\frac{dp_j}{dt} = p_j \left( G(v_j, \mathbf{L}) - G(E^t[v], \mathbf{L}) \right)$$

**does not** imply equation

$$\frac{dE^t[v]}{dt} \cong \sum_j v_j p_j \delta v_j \frac{\partial G(E^t[v], \mathbf{L})}{\partial v} = \sigma^2(t) \frac{\partial G(E^t[v], \mathbf{L})}{\partial v},$$

where $\sigma^2(t)$ is the variance of $v$ at time $t$.

Using the Taylor expansion up to the 2$^{nd}$ order, we get

$$G(v_j, \mathbf{L}) \cong G(E^t[v], \mathbf{L}) + \frac{\partial G(v, \mathbf{L})|_{v=u}}{\partial v} \delta v_j + \frac{1}{2} \frac{\partial^2 G(v, \mathbf{L})|_{v=u}}{\partial v^2} (\delta v_j)^2$$

where $\delta v_j = v_j - E^t[v]$, and therefore

$$E^t[G(v, \mathbf{L})] - G(E^t[v], \mathbf{L}) = \sum_j p_j \left( G(v_j, \mathbf{L}) - G(E^t[v], \mathbf{L}) \right) \cong \frac{1}{2} \sigma^2(t) \frac{\partial^2 G(v, \mathbf{L})|_{v=u}}{\partial v^2}.$$

Using the exact equation (2.14), we get an approximate equation:

$$\frac{dp_j}{dt} = p_j (G(v_j, \mathbf{L}) - E^t[G(v, \mathbf{L})]) =$$

$$p_j (G(v_j, \mathbf{L}) - G(E^t v, \mathbf{L}) + G(E^t v, \mathbf{L}) - E^t[G(v, \mathbf{L})]) \cong$$

$$p_j \left( \frac{\partial G(v, \mathbf{l})|_{v=u}}{\partial v} \delta v_j + G(E^t v, \mathbf{L}) - E^t[G(v, \mathbf{L})] \right) \cong$$

$$p_j \left( \frac{\partial G(v, \mathbf{L})|_{v=u}}{\partial v} \delta v_j - \frac{1}{2} \sigma^2(t) \frac{\partial^2 G(v, \mathbf{L})|_{v=u}}{\partial v^2} \right).$$

It follows from here that

$$\frac{dE^t[v]}{dt} \cong \sigma^2(t)\frac{\partial G(v,\mathbf{L})|_{v=u}}{\partial v} - \frac{1}{2}\sigma^2(t)\frac{\partial^2 G(v,\mathbf{L})|_{v=u}}{\partial v^2}. \tag{2.16}$$

Indeed,

$$\frac{dE^t[v]}{dt} = \sum_j v_j \frac{dp_j}{dt} \cong \sum_j v_j p_j \left(\frac{\partial G(v,\mathbf{L})|_{v=u}}{\partial v}\delta v_j - \frac{1}{2}\sigma^2(t)\frac{\partial^2 G(v,\mathbf{L})|_{v=u}}{\partial v^2}\right) =$$

$$\sum_j p_j \left(\frac{\partial G(v,\mathbf{L})|_{v=u}}{\partial v}(\delta v_j)^2 - \frac{1}{2}\sigma^2(t)\frac{\partial^2 G(v,\mathbf{L})|_{v=u}}{\partial v^2}\right) = \sigma^2(t)\frac{\partial G(v,\mathbf{L})|_{v=u}}{\partial v} - \frac{1}{2}\sigma^2(t)\frac{\partial^2 G(v,\mathbf{L})|_{v=u}}{\partial v^2}.$$

So, Equation (2.5b) may be a good approximation only if

$$\frac{\partial^2}{\partial v^2}G(v,\mathbf{L})|_{v=u} = 0,$$

i.e., if once again the function $G(v,\mathbf{L})$ is linear over the first argument.

Notice that according to the definition of covariance, we can write the (correct) equation (2.15) as

$$\frac{dE^t[v]}{dt} = \sum_j p_j v_j \big(G(v_j, \mathbf{L}) - E^t[G(v,\mathbf{L})]\big) = Cov^t[v, G(v,\mathbf{L})]. \tag{2.17}$$

One can recognize in the last equation the well-known covariance equation (Price, 1970, 1972; Robertson, 1968).

**Example 4.** Let again $\frac{dl_j(v)}{dt} = l_j v^2$, $G(v,\mathbf{v},\mathbf{L}) = v^2$. In this case equation (2.5b) becomes

$$\frac{dE^t[v]}{dt} = \sigma^2(t)\frac{\partial G}{\partial v}(E^t[v],\mathbf{L}) = 2\sigma^2(t)E^t[v] = 2(E^t[v^2]\,E^t[v] - (E^t[v])^3) = 2Cov^t[v,v^2].$$

On the other hand, the correct equation (2.17) becomes

$$\frac{dE^t[v]}{dt} = Cov^t[v, G(v,\mathbf{v},\mathbf{L})] = Cov^t[v, v^2].$$

This simple example shows that V&B equation (2.5b) unfortunately provides an incorrect approximation for the dynamics of the mean strategy in case of non-degenerate distribution of strategies and non-linear generating function $G$.

However, if $G$ is linear on the first argument, i.e. if $G(v,\mathbf{L}) = vG_1(\mathbf{L}) + G_2(\mathbf{L})$, then

$$Cov^t[v, G(v,\mathbf{v},\mathbf{L})] = Cov^t[v, vG_1(\mathbf{L}) + G_2(\mathbf{L})] = Var^t[v]G_1(\mathbf{L}) = \sigma^2(t)\frac{\partial G(v,\mathbf{L})}{\partial v}.$$

So far it has been shown that both equations of the model of Darwinian dynamics suggested by V&B, for which in the simplest case of a single species read

$$\frac{dx}{dt} = x\, G(E^t[v], \mathbf{v}, \mathbf{L}), \tag{2.18a}$$

$$\frac{dE^t[v]}{dt} = \sigma^2(t)\frac{\partial G}{\partial v}(E^t[v], \mathbf{u}, \mathbf{L}) \tag{2.18b}$$

cannot be derived from the underlying model (2.4) unless the function $G$ is linear in the first argument.

The correct equations for dynamics of the total species size and the mean value of strategies (traits) are well known in replicator dynamics, and in the case of a single species are given by equations (2.7) and (2.17):

$$\frac{dx}{dt} = x(t) E^t[G(v, \mathbf{L})], \tag{2.19a}$$

$$\frac{dE^t[v]}{dt} = Cov^t[v, G(v, \mathbf{L})]. \tag{2.19b}$$

Approximation of these equations up to the 2$^{nd}$ order are given as follows

$$\frac{dx}{dt} = x\left(G(u, \mathbf{L}) + \frac{1}{2}\sigma^2(t)\frac{\partial^2}{\partial v^2} G(v, \mathbf{L})|_{v=u}\right), \tag{2.20}$$

$$\frac{du}{dt} = \sigma^2(t)\frac{\partial G(v,\mathbf{L})|_{v=u}}{\partial v} - \frac{1}{2}\sigma^2(t)\frac{\partial^2 G(v,\mathbf{L})|_{v=u}}{\partial v^2}.$$

One can see that the V&B model (2.18) is the 1$^{st}$ order approximation of the exact model (2.19a,b), and so model (2.18) can be used only if $\sigma^2(t)\frac{\partial^2}{\partial v^2} G(v, \mathbf{L})|_{v=u} \cong 0$ for all $t$.

Similarly, both equations of the general model of Darwinian dynamics formulated by V&B in Chapter 5 as equations (5.1) and (5.9) for multi-species populations do not follow from the initial dynamical system (5.2) for phenotype dynamics unless the distribution of strategies is concentrated in a single point, or if the generating $G$-function is linear in the first argument, significantly reducing the scope of applicability of those equations.

**3. The model of "Darwinian dynamics" contains an internal inconsistency**

It was shown in the previous section that the model of Darwinian dynamics, as it was formulated by V&B, cannot be derived in the general case from the underlying dynamics of phenotypes that compose the species and the population.

Nevertheless, one can pose the following question: is it possible to ignore the (perhaps, unknown) dynamics of phenotypes and just to assume the existence of such G-function that equations (2.18a,b) hold for the density of inhomogeneous species and the mean value of strategies within the species?

It seems that V&B (p.121, Eq 5.1)) considered such a possibility as evident (page 121):

> "We refer to a **species** as a set of evolutionarily identical individuals whose strategies, referred to as **phenotypes**, aggregate around a distinct *mean strategy value*…We show below that, when the phenotypes of a particular species aggregate around some mean strategy value (interbreeding may facilitate this aggregation), the distribution of strategies results in an evolutionary dynamic. In this case… the variable $u_i$ no longer refers to the strategy of any given individual in the population $x_i$, but rather we now define $u_i$ to be the **mean strategy** of all individuals in the population. "

> The correctness of this approach is not evident and may be related with the initial conceptions of the evolutionary game. According to (McGill & Brown, 2007), "A key innovation of evolutionary game theory involves extending the classical notion of facing a single opponent playing strategy $U$ to facing a population playing strategy $U$. In Maynard Smith's (1982, p. 23) idea of "playing the field," the individual does not interact in a pair-wise fashion with other individuals; rather the individual faces an opponent that is the population at large. In this case we can think of the entire population playing the single strategy $U$. Under this interpretation we see $u$ as the strategy of a mutant individual or a focal individual, and $U$ is the resident strategy. Even if individuals in the resident population show variation, the playing-the-field approach can still work if the fitness of the target individual is well approximated by considering the average strategy of the resident population".

Of course, the approach based on the idea of "playing the field" can work if indeed "the fitness of the target individual is well approximated by considering the average strategy of the resident population". The question is: under what conditions does this assumption hold? In what follows I show that the necessary conditions are unfortunately quite restrictive.

Let us again consider for simplicity the case of one-species many-strategies Darwinian model that has a form

$$dx/dt = xG(u,x) \qquad (3.1a)$$

$$\frac{du}{dt} = \sigma^2 \frac{\partial G(u,x)}{\partial u}, \qquad (3.1b)$$

where $x$ is the population size and $u = E^t v$ is the mean strategy at $t$ moment. The approach based on the assumption that the corresponding G-function exists, results in a system of only two equations; it gives an attractive way to transition from describing individual dynamics to describing dynamics of a species as a whole. This approach was used, e.g., in application to a logistic model in (R. C. Rael, 2009) and the Allee-type model in (Cushing & Hudson, 2012; R. Rael et al., 2009); see also V&B for many other models. For example, in case of one-species many-strategies Darwinian logistic model the G-function used in the cited papers has a form

$$G(v,x) = r(v)\left(1 - \frac{x}{K(v)}\right). \qquad (3.2)$$

Let us emphasize that in model (3.1a,b) $u$ is **the mean value** of strategies within an inhomogeneous population, where different individuals can use different strategies. Neither initial, nor current distributions of strategies within the population are specified in model (3.1a,b). Hence, it was implicitly assumed that equations (3.1a,b) hold for any initial distribution of strategies.

Nevertheless, V&B pointed out in their derivation of equations (2.5a,b) that "some thought also needs to be given to how the original population is distributed among the phenotypes. A reasonable assumption would be to assign the majority of the population to phenotypes in the neighborhood of the mean strategy."

Below I show that this assumption is not sufficient, and that in particular, arbitrarily small initial variance can become arbitrarily large over time.

Let us emphasize that in order to define the mean strategy in the population at $t$ moment using equation (3.1b), one needs to know the variance of the current distribution of strategies within the population at this moment, which in general is unknown. To overcome this difficulty one can just use the definition of the mean value.

We do not want to restrict ourselves to the case when the set $U$ of possible strategies is finite and the strategy distribution is discrete. Denote again $x(t)$ as the population size and $l(t, v)$ as the density at $t$ moment of a phenotype that uses strategy $v$, so that $x(t) = \int_U l(t,v)dv$. Then the distribution of strategies is given by

$$P(t,v) = l(t,v)/x(t), \text{ and } u = E^t[v] = \int_U vP(t,v)dv.$$

The definition of a G-function (V&B, p.121 Eq.(5.1)) associates the fitness of a species at each time moment with the mean strategy of the species at this moment, $E^t[v]$, see equations (2.5a). Let us emphasize that according to this definition, the *initial* distribution of strategies can be *arbitrary*, and the dynamics of the distribution is determined by the dynamics of the species and its composition.

At a first glance, the only problem with model (3.1) is that it cannot be derived from the underlying dynamics of phenotypes, which compose the population (see the previous section). Is it possible just to assume the existence of such a G-function, so that equations (3.1a,b) hold for the density of inhomogeneous species and the mean value of strategies within the species? Assume the density of individuals using strategy $v$ is given by equation

$$\frac{dl(t,v)}{dt} = l(t,v)H(v,x). \tag{3.3}$$

Let us assume that there exists a function $G$, such that equation (3.1a) holds. It follows from equation (3.3) that

$$\frac{dx}{dt} = xE^t[H(v,x)].$$

Then, the function $G(u,x)$ should satisfy the following equation:

$$G(E^t[v], x) = E^t[H(v,x)],$$

i.e.,

$$G(\int_V vP(t,v)dv, x) = \int_V H(v,x)P(t,v)dv$$

at any initial distribution.

Assume that the initial distribution $P(0,v)$ is concentrated in (or around) point $v_0$. Then $G(v_0, x) = H(v_0, x)$ for any $v_0 \in V$. However, we have already proven that it is possible only if the function $H(v,x)$ is linear over $v$. Therefore, in general it is impossible to define a function $G(v,u,x)$ such that the Darwinian model (3.1a,b) would be correct for an arbitrary initial distribution. Consequently, this model may be used only under serious restrictions. Moreover, it means that the concept of "playing the field" may be wrong and contain internal contradictions unless the fitness function is linear.

## 4. The case of linear fitness function

It was previously emphasized that equations of Darwinian model (2.5a,b) are correct if and only if the fitness function $H$ is linear over the first argument,

$$H(v,x) = f(x) + vg(x). \tag{4.1}$$

In this case the model becomes:

$$\frac{dl(t,v)}{dt} = l(t,v)H(v,x) = l(t,v)[f(x) + vg(x)], \tag{4.2}$$

$$x(t) = \int_V l(t,v)dv.$$

The following equations follow from (4.2) and (2.17):

$$\frac{dx}{dt} = x(f(x) + E^t[v]g(x)) = xH(E^t[v], x), \tag{4.3}$$

$$\frac{du}{dt} = \frac{dE^t[v]}{dt} = Cov[v, f(x) + vg(x)] = g(x)Var^t[v] = \sigma^2(t)\frac{\partial}{\partial v}H(v,x). \tag{4.4}$$

Equations (4.3) - (4.4) coincide with the model of Darwinian evolution of V&B (3.1 a, b), and this model is correct only when the fitness function $H(v,x)$ is linear over $v$. I have considered only a simple situation with a single species and scalar strategies, but in more general cases of many species and vector strategies the situation is similar.

Notice that equation (4.4),

$$\frac{dE^t[v]}{dt} = Cov[v, f(x) + vg(x)] = g(x)Var^t[v],$$

is a version of Fisher' Fundamental theorem; more generally, this equation is a special case of the Price' equation. It is well known that the Price' equation is "dynamically insufficient", i.e., it cannot be used *alone* as a propagator of model dynamics over time. The problem was discussed in (Barton & Turelli, 1987; Frank, 1997; Lewontin, 1974; Rice, 2006).

The essence of the problem is that in order to compute the mean $E^t[u]$ in equation (4.4), one needs to know the variance $\sigma^2(t)$; in turn, to compute the variance using Price' equation, one needs to know the 3$^{rd}$ order moment, and so on. For this reason, model (4.3), (4.4) is also "dynamically insufficient". Nevertheless, model (4.3), (4.4) can be explicitly solved if the initial distribution of $v$ is known, using the HKV (hidden keystone variables) method, see (Kareva & Karev, 2019). As for the general model, the current variance $\sigma^2(t)$ is unknown and hence the model cannot be solved without additional assumptions.

In the paper (R. C. Rael, 2009) and some other papers (see also some examples in V&B), the authors just assumed that $\sigma^2(t)$ is a constant. Let us emphasize that the variance $\sigma^2(t)$ is defined by the current distribution of strategies $v$, which is determined by the system dynamics and initial distribution of strategies; it evolves over time and cannot be assigned arbitrarily. Notably, $\sigma^2(t)$ can be constant only in some special cases (see s.5 below), in contrast to the statement made in s.5.12 of V&B.

An important example of the Darwinian model for logistic population growth was considered in V&B, s. 5.1 and in (R. C. Rael, 2009).

Assume a population is composed of species that grow logistically; assume that the Malthusian growth rate and the carrying capacity of every species depend on its strategy. Then the dynamics of the species is described by the following equation:

$$\frac{dl(v,t)}{dt} = l(v,t)r(v)\left(1 - \frac{x}{K(v)}\right),$$

where $x(t)$ is the total population size,

$$x(t) = \int_V l(v,t)dv$$

and where $V$ is the set of all possible strategies. The current distribution of strategies is defined as

$$P(v,t) = l(v,t)/x(t).$$

Initial distribution $P(0,t)$ is assumed to be known.

V&B define the G-function for this model as

$$G(v,x) = r(v)\left(1 - \frac{x}{K(v)}\right),$$

where
$$K(v) = K\exp(-\tfrac{v^2}{2\sigma^2}).$$

Then, the V&B equations become

$$\frac{dx}{dt} = xG(u,x) = x\, r(u)\left(1 - \frac{x}{K}\exp\left(\frac{u^2}{2\sigma_K^2}\right)\right) \tag{4.5a}$$

$$\frac{du}{dt} = \sigma^2 \frac{\partial G(u,x)}{\partial u} = \sigma^2 \frac{\partial}{\partial u} r(u)\left(1 - \frac{x}{K}\exp\left(\frac{u^2}{2\sigma_K^2}\right)\right). \tag{4.5b}$$

Unfortunately, the first equation in general may be incorrect. Assume for simplicity that $r(v) = r = const$. Equation for phenotype dynamics (2.4) of a species with *individual strategy v* can be written in current notation as

$$\frac{dl(v,t)}{dt} = l(v,t) r\left(1 - \frac{x}{K(v)}\right) = l(v,t) r\left(1 - \frac{x(t)}{K}\exp\left(\frac{v^2}{2\sigma^2}\right)\right).$$

Then, integrating the last equation over $v$ results in the following equation:

$$\frac{dx}{dt} = x(t)(r - \frac{x(t)}{K}\int_V \exp\left(\frac{v^2}{2\sigma^2}\right) P(t,v) dv = x(t)(r - \frac{x(t)}{K} E^t\left[\exp\left(\frac{v^2}{2\sigma^2}\right)\right].$$

In general

$$E^t\left[\exp\left(\frac{v^2}{2\sigma^2}\right)\right] \neq \exp(E^t[v]^2/2\sigma^2)$$

and so

$$\frac{dx}{dt} \neq x\, G(E^t[v], x).$$

Notice that in the case when $r(v) = r = const$ and $\sigma^2 = const$, Equations (4.5a,b) have a first integral; indeed,

$$\frac{du}{dt} = \sigma^2 \frac{\partial}{\partial u} r\left(1 - \frac{x}{K}\exp\left(\frac{u^2}{2\sigma_K^2}\right)\right) = -\sigma^2 \frac{rx}{K}\exp\left(\frac{u^2}{2\sigma_K^2}\right) u/\sigma_K^2,$$

$$\frac{dx}{dt} = x\left(r - \frac{rx}{K}\exp\left(\frac{u^2}{2\sigma_K^2}\right)\right) = x(r + \sigma_K^2/\sigma^2 \frac{du}{dt}/u), \text{ or}$$

$$\frac{d\ln x}{dt} = r + \sigma_K^2/\sigma^2 \frac{d\ln u}{dt}.$$

Therefore,

$$\ln x = rt + \sigma_K^2/\sigma^2 \ln u + C = rt + \ln u + \ln x(0) - \sigma_K^2/\sigma^2 \ln u(0), \text{ and}$$

$$x(t) = \frac{x(0)}{u(0)^{\sigma_K^2/\sigma^2}} u(t)\exp(rt). \tag{4.6}$$

Using this equation, one can reduce System (4.5) to a single equation for the mean:

$$\frac{du}{dt} = -\frac{\sigma^2}{\sigma_K^2}\frac{rx}{K}\exp\left(\frac{u^2}{2\sigma_K^2}\right)u = -C\,u^2\exp\left(\frac{u^2}{2\sigma_K^2}+rt\right),\qquad(4.7)$$

where

$$C = \frac{\sigma^2}{\sigma_K^2}\frac{r}{K}\frac{x(0)}{u(0)\frac{\sigma_K^2}{\sigma^2}}$$

and then compute $x(t)$ using Equation (4.6).

An example is given below.

**Example 5.** Solution to equations (4.6), (4.7):

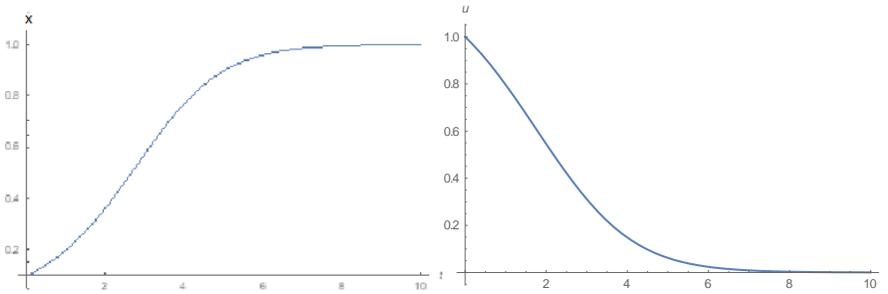

**Figure 1**. Solution to Equations (4.6), (4.7) when $u(0) = 1, x(0) = 0.1, r = 1, \sigma_K = 1, \frac{\sigma^2}{K} = 1$; population size (left); mean value of the trait (right).

According to Equation (4.7), $\frac{du}{dt} < 0$ while $\sigma^2 > 0$ and $u \neq 0$. If the domain of $v$ is $[a, b), a \geq 0, b \leq \infty$, then $u(t) = E^t[v] \to a$, so the distribution of strategies concentrates over time in the point $a$, and consequently variance $\sigma^2 \to 0$. Hence, the assumption that $\sigma^2 = const$ **cannot be valid** for any initial distribution concentrated in $[a, b)$.

The Darwinian logistics model with distributed Malthusian rate and carrying capacity was studied in (R. C. Rael, 2009). The initial logistic model was given by the equation

$$\frac{dl(v,t)}{dt} = l(v,t)G(x,v) = l(v,t)r(v)\left(1 - \frac{x(t)}{K(v)}\right),\qquad(4.8)$$

where $l(v, t)$ is the species density, $x(t)$ is the total population at time $t$, and

$$K(v) = K_m exp\left(-\frac{(v-k)^2}{2\sigma_k^2}\right),$$

$$r(v) = r_m exp\left(-\frac{(v-\rho)^2}{2\sigma_r^2}\right).$$

The Darwinian version of this model has the form

$$\frac{dx}{dt} = xr_m exp\left(-\frac{(u-\rho)^2}{2\sigma_r^2}\right)\left(1 - \frac{x}{K}exp\left(\frac{(u-k)^2}{2\sigma_K^2}\right)\right), \quad (4.9)$$

$$\frac{du}{dt} = \sigma^2 \frac{\partial}{\partial u}\left(r_m exp\left(-\frac{(u-\rho)^2}{2\sigma_r^2}\right)\left(1 - \frac{x}{K}exp\left(\frac{(u-k)^2}{2\sigma_K^2}\right)\right)\right).$$

A typical solution to this equation is shown on Fig.2

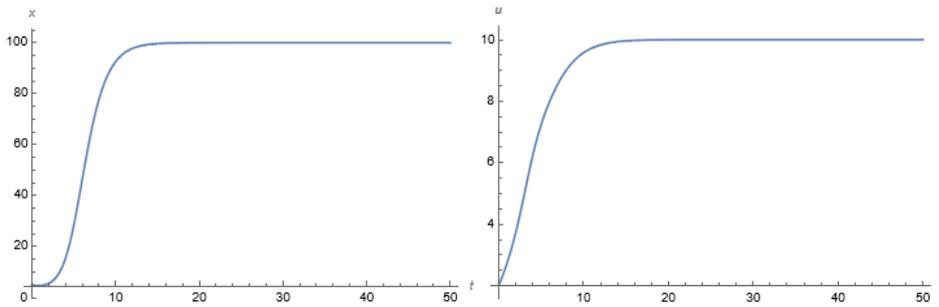

**Figure 2**. Solution to Darwinian logistic model (4.9) with $r_m = 1, \rho = 7, \sigma_r^2 = 20, {\sigma_K}^2 = 18, K = 100, k = 10$; population size (left); mean value of the trait (right).

Notably, the model (4.9) does not take into account the range of possible values or the initial distribution of the traits (or strategies), and instead considers only the mean value of the trait. Below I show that different initial distributions may result in qualitatively different population dynamics.

First consider a discrete initial distribution. Notice that in reality, any population consists in a finite number of species; formally, any distribution may be approximated by a discrete one. Therefore, assume that the trait can take the values $\{v_i, i = 0, ... n\}$. Let $x_i(t)$ be the size of $i$-th species at $t$ time. As an example, let $n = 10$ and $x_i(0) = 1$ $v_i = i, i = 0, \ldots n$. The following Fig.3 shows dynamics of population size and mean of the trait in this case as they change over time.

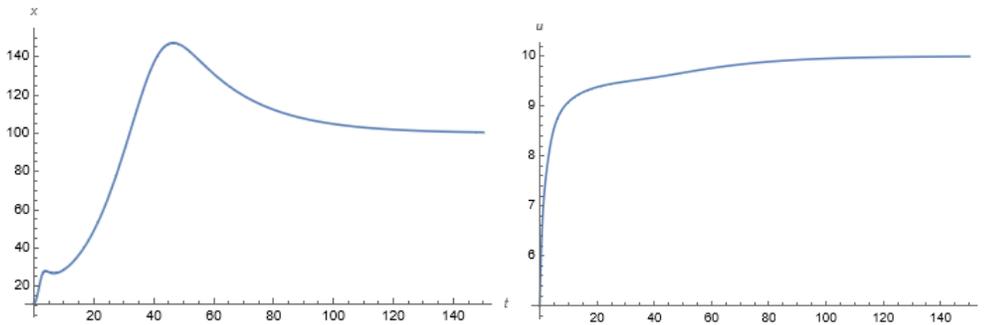

**Figure 3**. Solution to inhomogeneous logistic model (4.8) with discrete initial distribution of the trait at $l_i(0) = 1$, $v_i = i, i = 0, \ldots 10$ with $r_m = 1, \rho = 7, \sigma_r^2 = 20, \sigma_K^2 = 18, K = 100, k = 10$; population size (left) and mean value of the trait (right).

It is interesting to trace the dynamics of individual species, given by equation (4.8) at $l_i(0) = 1$, $v_i = i, i = 0, \ldots 10$, see Figures 4 and 5; the parameter values are given in the legend for Fig.3.

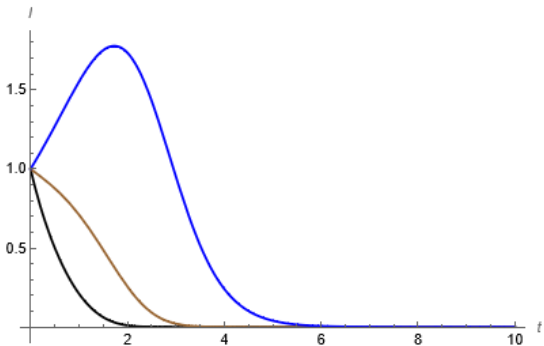

**Figure 4.** Plots of the species $l_1$(black), $l_3$ (brown) and $l_5$ (blue). Parameter values are given in the legend for Figure 3.

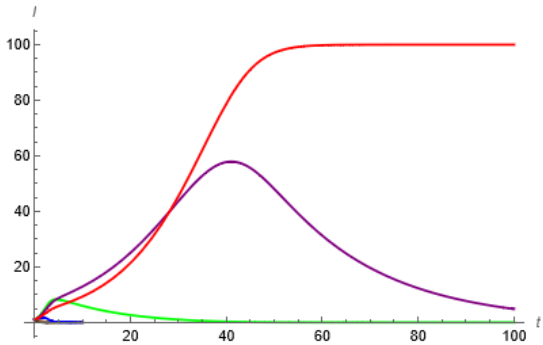

**Figure 5**. Plots of the species $l_8$ (green), $l_9$(purple), and $l_{10}$ (red). Parameter values are given in the legend for Figure 3.

One can see that all species become extinct over time except for the last species $x_{10}$.

Now consider a population given by equation (4.8) at $l_i(0) = \exp(\frac{1-0.1i}{2})$, $v_i = i/10$, $i = 0, \ldots 10$; the parameter values are given in the legend for Fig.3. The following Fig.6 shows the dynamics of the population size and mean trait in this case; Figures 7 and 8 show the dynamics of individual species.

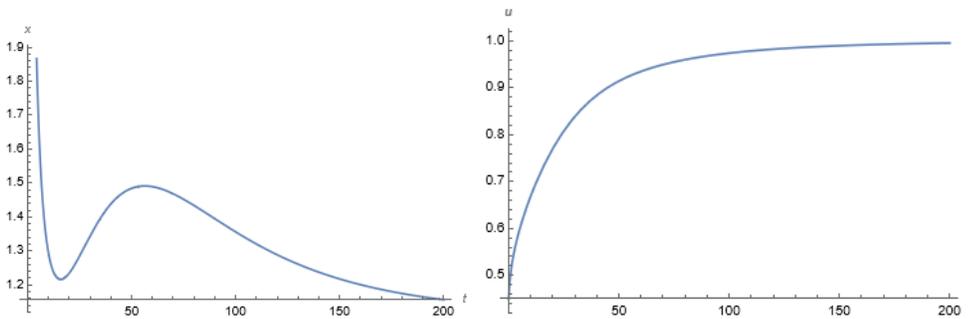

**Figure 6**. Solution to inhomogeneous logistic model (4.8) at $l_i(0) = \exp(\frac{1-0.1i}{2})$ and $v_i = i/10$, $i = 0, \ldots 10$; population size (left), mean value of the evolving trait (right); other parameter values are given it the legend for Figure 3.

As one can see in Figs. 7 and 8, the dynamics of individual species in this case are qualitatively different from the previous one.

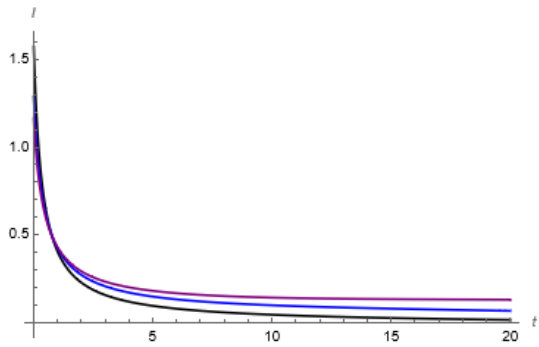

**Figure 7**. Plots of species $l_1$ (black), $l_5$ (blue), $l_7$ (purple) given by logistic model (4.8) at $v_i = i/10$ and $l_i(0) = \exp(\frac{1-0.1i}{2})$, $i = 0, \ldots 10$. Parameter values are given in the legend for Figure 3.

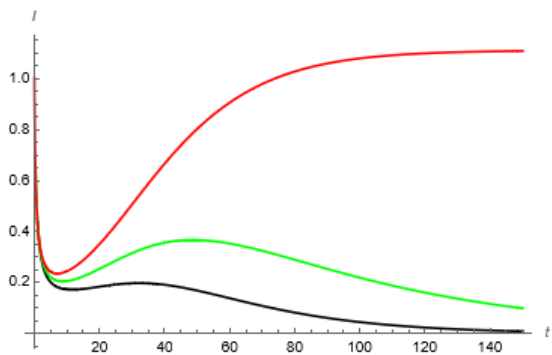

**Figure 8**. Plots of species $l_8$ (black), $l_9$ (green), $l_{10}$ (red) given by logistic model (4.8) at $v_i = i/10$ and $l_i(0) = \exp(\frac{1-0.1i}{2})$, $i = 0, \ldots 10$. Parameter values are given in the legend for Figure 3.

Overall, Figures 3 and 6, which describe change over time in population size and the mean trait for two different sample initial distributions, show that these dynamics very much depend on the domain of the trait, its initial distribution, and the initial distribution of species numbers, and can differ dramatically from the dynamics predicted by the Darwinian logistic model (4.9). Ignoring these characteristics can result in qualitatively wrong predictions.

**Example 6.** Consider the inhomogeneous model of the form
$$\frac{dl(v,t)}{dt} = l(v,t)(1 - v^2 x), \; v \in [0,1] \tag{4.10}$$

where $x(t) = \int_0^1 l(v,t) dv$ is the population size. G-function for this model is
$$G(v,x) = (1 - v^2 x).$$
The Darwinian model in this case reads as follows:
$$\frac{dx}{dt} = xG(u,x) = x(1 - u^2 x) \tag{4.11}$$
$$\frac{du}{dt} = \sigma^2 \frac{\partial G(u,x)}{\partial u} = -2u\sigma^2 x.$$
One cannot solve this system directly since the current variance $\sigma^2(t)$ is unknown. If one were to assume that $\sigma^2(t) = \sigma^2$ is a constant, then the solution to the system (4.11) is shown on Fig.7:

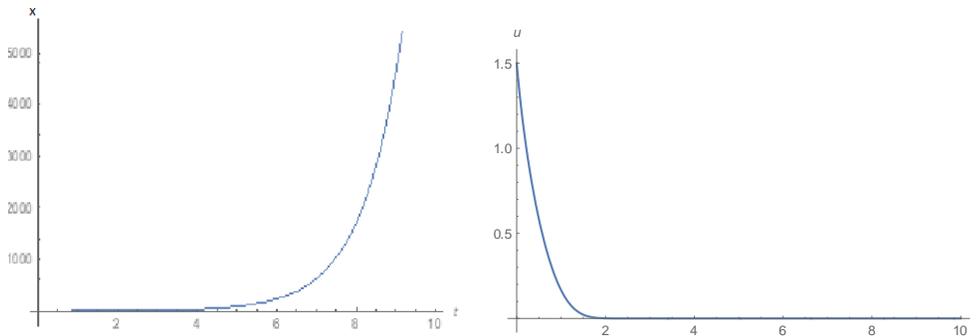

**Figure 7**. Solution to System (4.11) for $\sigma^2 = 1, x(0) = 1, u(0) = 1.5$. Population size is shown on the left; mean of the trait is shown on the right.

On the other hand, the model (4.10) can be solved explicitly using the HKV method (Kareva & Karev, 2019). Introduce an auxiliary variable

$$\frac{dq}{dt} = x, q(0) = 0.$$

Then

$$\frac{dl(v,t)}{dt} = l(v,t)\left(1 - v^2 \frac{dq}{dt}\right)$$

and

$$l(v,t) = l(v,0)\exp(t - v^2 q(t)).$$

Next, it is necessary to choose the initial distribution of strategies; let, for example, $v \in [1,2]$ and let the initial distribution of $v$ be uniform. Then

$$x(t) = x(0) \int_1^2 e^{t-v^2 q(t)} dv = x(0) e^t \sqrt{\frac{\pi}{q(t)}} \frac{Erf(\sqrt{2q(t)}) - Erf(\sqrt{q(t)})}{2}. \quad (4.12)$$

Hence, an equation for the variable $q(t)$ is:

$$\frac{dq}{dt} = x(0) e^t \sqrt{\frac{\pi}{q(t)}} \frac{Erf(\sqrt{2q(t)}) - Erf(\sqrt{q(t)})}{2}. \quad (4.13)$$

This equation can be solved numerically. With this solution, it is possible to compute the population size (4.12) and all statistical characteristics of the model, such as the moment generating function (mgf) of the current distribution, mean and variance at any time using the following equations:

$$P(t,v) = \frac{l(v,t)}{x(t)} = 2\sqrt{\frac{q(t)}{\pi}} e^{-v^2 q(t)} / (Erf(\sqrt{2q(t)}) - Erf(\sqrt{q(t)})); \quad (4.14)$$

$$u(t) = E^t[v] = \frac{e^{-q(t)} - e^{-4q(t)}}{\sqrt{\pi q(t)}(Erf(2\sqrt{q(t)}) - Erf(\sqrt{q(t)}))}; \quad (4.15)$$

$$\sigma^2(t) = \frac{1}{2q(t)} + \frac{e^{-q(t)} - 2e^{-4q(t)}}{\sqrt{\pi q(t)}(Erf(\sqrt{2q(t)}) - Erf(\sqrt{q(t)}))} - \frac{(e^{-q(t)} - e^{-4q(t)})^2}{\pi q(t)(Erf(\sqrt{2q(t)}) - Erf(\sqrt{q(t)}))^2}. \quad (4.16)$$

The following figures 8-10 show the dynamics of total population size, the mean, variance and distribution of traits given by the exact solution to the model (4.10).

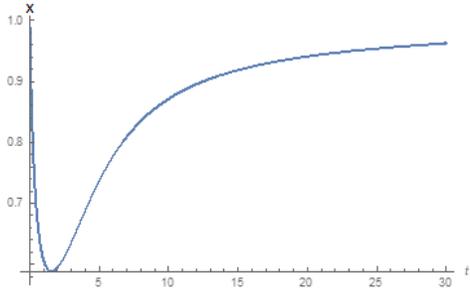

**Figure 8**. Exact solution of the model (4.10); initial distribution is uniform in $[1,2]$, $x(0) = 1$.

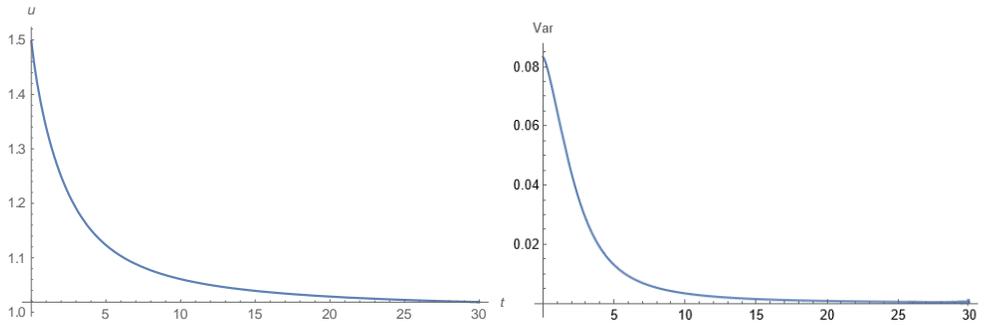

**Figure 9**. Dynamics of the mean (left) and variance (right) given by equations (4.15) and (4.16) for model (4.10).

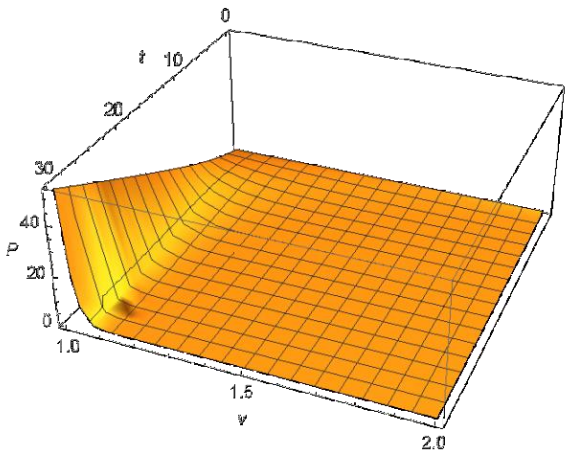

**Figure 10**. Dynamics of the current pdf of model (4.10), see Eq. (4.14). Initial uniform distribution in
[1, 2] tends over time to a distribution concentrated in the minimal possible value $v = 1$.

One can see the dramatic difference between the exact solution of model (4.10) and the solution to Darwinian V&B model (4.11) under assumption that $\sigma^2 = 1$. One can see also that the variance of the strategy distribution for model (4.10) in no case can be assumed to be a constant.

Overall, these results have shown that in the case of a linear fitness, the V&B model (at least, the first equation) is correct, and the underlying model of species dynamics can be solved for an arbitrary initial distribution. In case of non-linear fitness function the equation for total population size as given by Darwinian model cannot be derived from the underlying species dynamics and can (and as a rule, does) predict wrong dynamics of population size. Let us emphasize that in general the assumption $\sigma^2(t) = const$ does not hold even in the case of linear fitness function and hence the 2nd equation of the V&B model may be wrong. In the next section I show that the only case when $\sigma^2(t) = const$ is the normal initial distribution. In case of other initial distributions this assumption may result dramatically wrong predicted dynamics of the model.

**5. Variance dynamics**

In the previous section it was shown that the assumption $\sigma^2(t) = const$ may be wrong and can predict incorrect dynamics of the model characteristics. To better understand this issue, consider the section V&B 5.12 "Variance dynamics" in detail. In this section, the authors tried to show that variance does not change over time under the assumption of a symmetric distribution of strategies and small (initial) variance, and that variance becomes smaller as the system approaches equilibrium (notice that these two statements contradict each other). It will be shown next that the derivation of these statements contains a mistake and both statements can unfortunately be incorrect.

To better understand the computations in V&B, s.5.12, consider again the simplified case of a single population composed of different species (so index $i$ is omitted in all formulas of s.5.12). Then, by definition,

$$\sigma^2(t) = \sum_j (\delta v_j)^2 q_j(t)$$

where $\delta v_j = v_j - E^t[v]$ and $q_j(t)$ is the frequency of species $j$ at time $t$. Therefore,

$$\frac{d\sigma^2(t)}{dt} = \sum_j \left[\frac{d(\delta v_j)^2}{dt} q_j + (\delta v_j)^2 \frac{dq_j}{dt}\right] = \sum_j (\delta v_j)^2 \frac{dq_j}{dt}$$

as

$$\sum_j \frac{d(\delta v_j)^2}{dt} q_j = -2 \frac{dE^t[v]}{dt} \sum_j (u_j - E^t[v]) q_j = 2 \frac{dE^t[v]}{dt} (\sum_j v_j q_j - E^t[v]) = 0.$$

Next, an elementary calculus mistake was done in the *incorrect* equation in V&B (5.25):

$$\dot{q}_J = \frac{\dot{x}_J}{\dot{x}},$$

while the correct equation should be

$$\dot{q}_J = \frac{d}{dt}\left(\frac{x_J}{x}\right) = \frac{x\dot{x}_J - x_J\dot{x}}{x^2},$$

and hence all further computations on p.150 of V&B, which result the final equation $\frac{d\sigma^2}{dt} = 0$ are unfortunately irrelevant. Noticeably, the correct formula $\dot{q}_J = \frac{d}{dt}\left(\frac{x_J}{x}\right) = \frac{x\dot{x}_J - x_J\dot{x}}{x^2}$ was used previously on p.123; it easily implies that

$$\dot{q}_J = \frac{\dot{x}_J}{x} - q_j \frac{\dot{x}}{x} = q_j(H_j - E^t[H]),$$

which is a standard replicator equation.

The correct equation for the variance is well-known and follows directly from the Price' equation (Price, 1970) that holds for any random variable $z_t$ such that $\frac{dz}{dt} = zF$:

$$\frac{d}{dt} E[z_t] = Cov[z_t, F] + E\left[\frac{dz_t}{dt}\right].$$

Taking $z_t = (v - E^t v)^2$, we get

$$\frac{d\sigma^2(t)}{dt} = \frac{d}{dt} E^t[(v - E^t[v])^2] = Cov[H(v,x), (v - E^t[v])^2] - 2E^t[(v - E^t[v]) \frac{dE^t[v]}{dt}];$$

$$E^t\left[(v - E^t[v]) \frac{dE^t[v]}{dt}\right] = \frac{dE^t[v]}{dt} E^t[u - E^t[v]] = 0,$$

and therefore

$$\frac{d\sigma^2(t)}{dt} = Cov[H(v,x), (v - E^t[v])^2]. \tag{5.1}$$

Hence, $\frac{d\sigma^2}{dt} = 0$ only if

$$Cov[H(v,x),(v-E^t[v])^2] = 0,$$

i.e. if

$$E^t[H(v,x)(v-E^t[v])^2] = E^t[H(v,x)]E^t[(v-E^tv)^2] = \sigma^2(t)\,E^t[H(v,x)].$$

In order to make the problem of variance dynamics more transparent, consider the simplest case of an inhomogeneous Malthusian population

$$\frac{dl(t,v)}{dt} = vl(t,v), \qquad (5.2)$$

with the G-function $G(v,\mathbf{u}) = v$.

The equations of Darwinian dynamics are valid for this model:

$$\frac{dx}{dt} = xE^t[v], \qquad (5.3a)$$

$$\frac{dE^t[v]}{dt} = \sigma^2(t), \qquad (5.3b)$$

where total population size $x(t) = \int_V l(t,v)dv$.

Once again, the system (5.3a,b) is not closed as $\sigma^2(t)$ is unknown and critically depends on the distribution of strategies $v$ and its dynamics.

One can easily overcome this problem in case of Malthusian model (5.2). Indeed, equation (5.2) has a solution

$$l(t,v) = l(0,v)exp(vt).$$

and so

$$x(t) = \int_V l(0,v)exp(vt)dv = x(0)(\int_V P(0,v)exp(vt)dv = x(0)M_0(t), \qquad (5.4)$$

where $M_0(\delta) = \int_V exp(v\delta)P(0,v)dv$ is the moment generating function (mgf) of the initial distribution of $v$.

The current distribution is given by

$$P(t,v) = \frac{l(t,v)}{x(t)} = \frac{exp(vt)}{M_0(t)}P(0,v), \qquad (5.5a)$$

the mean value is given by

$$E^t[v] = \frac{d}{dt}M_0(t)/M_0(t), \tag{5.5b}$$

and the variance is

$$\sigma^2(t) = \frac{d^2 M_0(t)}{dt^2}/M_0(t) - (E^t[v])^2. \tag{5.5c}$$

The following equation for the mgf $M_t(\lambda) = \int_V exp(v\lambda)P(t,v)dv$ of the distribution at $t$ time immediately follows from equation (5.5a); it allows us to trace the evolution of the distribution over time:

$$M_t(\lambda) = M_0(\lambda + t)/M_0(t). \tag{5.6}$$

Assume that the initial distribution of the strategies is normal with the mean $m_0$, variance $\sigma_0^2$, and mgf $M_0[\lambda] = e^{\frac{\lambda^2 \sigma_0^2}{2} + \lambda m_0}$. Then, using equation (5.6), it is easy to show that the strategy distribution will also be normal at any time $t$ with the mean $m_t = m_0 + t\sigma_0^2$ and with the same variance $\sigma_0^2$. Indeed, according to equation (5.6),

$$M_t[\lambda] = \frac{M_0[\lambda+t]}{M_0[t]} = e^{\left(\frac{\lambda^2 \sigma_0^2}{2} + \lambda(m_0 + t\sigma_0^2)\right)}.$$

It is exactly the mgf of the normal distribution with the mean $E^t[v] = m_0 + \sigma_0^2 t$ and variance $\sigma_0^2$. So, in the case of a normal distribution, the assumption that the variance of strategy distribution is a constant is justified (below I show that the normal distribution is the only case when the variance remains constant over time).

In contrast, assume that the initial distribution of strategies is *exponential*, i.e., $P(0,v) = e^{-sv}/s$ with parameter $s$; its mean is $E^0[v] = 1/s$, variance $\sigma^2(0) = \frac{1}{s^2} = E^0[v]^2$ and the mgf $M_0[\lambda] = \frac{s}{s-\lambda}$. In this case, according to the equation (5.6), the mgf of the current distribution is

$$M_t[\lambda] = \frac{M_0[\lambda+t]}{M_0[t]} = \frac{s-t}{s-t-\lambda};$$

it is the mgf of the exponential distribution $P(t,v) = e^{-(s-t)v}/(s-t)$ with parameter $s - t$. The population increases in such a way that the distribution of strategies is exponential at every instant $t < s$ with the mean $E^t[v] = 1/(s-t)$ and the variance $\sigma^2(t) = 1/(s-t)^2$. So, the variance can be arbitrarily small ($= 1/s^2$) at the initial time but will increase indefinitely as $t \to s$ together with the mean value of the strategy parameter and the total population size.

Next, let us consider exponential initial distribution truncated on the interval $[0, b]$, where $P(0, v) = Ce^{-sv}, v \in [0, b]$, $C = s/(1 - e^{-bs})$, $s$=const.
Then, according to equation (5.5a), the current distribution at each time moment is
$$P(t, v) = \frac{e^{-(s-t)v}(s-t)}{1 - e^{-(s-t)b}}, t < s.$$
It is again truncated exponential distribution with parameter $s - t$.

Its mean is $E^t[v] = \frac{1}{s-t} - \frac{b}{e^{b(s-t)} - 1}$ and its variance is $\sigma^2(t) = \frac{1}{(s-t)^2} - \frac{b^2 e^{b(s+t)}}{(e^{bs} - e^{bt})^2}$.

Notice that $\lim_{t \to \infty} E^t[v] = b$ and $\lim_{t \to \infty} \sigma^2(t) = 0$. This means that over time, the distribution will concentrate in point $b$, and the variance will vanish, leaving a single clone.
All these three possibilities are shown in Figure 11.

Notice that for model (5.2), for any initial distribution concentrated on a bounded interval $[a, b]$, $\frac{dE^t[v]}{dt} = \sigma^2(t) > 0$. Therefore, the mean value increases until the distribution is not concentrated in a single point. This means that over time, the distribution will tend to the distribution concentrated in the point $b$ and the variance will tend to 0.

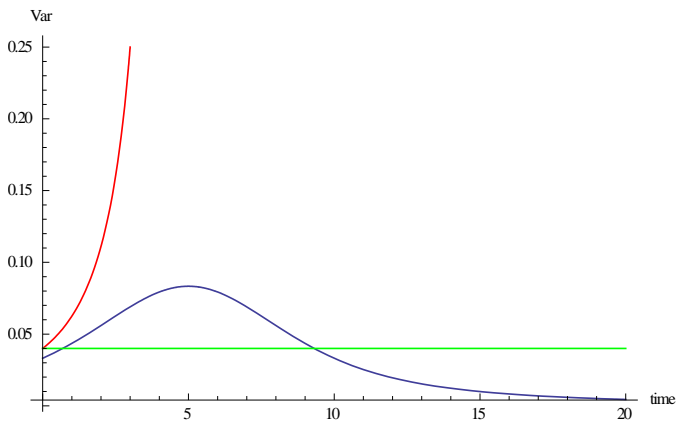

**Figure 11**. Dynamics of the variance of Malthusian model (5.2) depending on initial distribution.

a) normal distribution (green); b) exponential distribution with the mean 1 (red); c) truncated exponential distribution in [0,5] (blue).

Next, consider the claim that the variance is constant if the distribution of strategies is symmetric, and variance is small (see V&B, section 5.12). The following example shows that this statement is not valid if these conditions hold only for the initial distribution.

**Example 7.** Consider the Malthusian inhomogeneous model (5.2) and assume that the initial distribution is double exponential (also known as the Laplace distribution), given by the formula

$$P(0,v) = \frac{s}{2}\exp(-s|v-m|).$$

Notice that this distribution is *symmetric* around the mean value $m$ and its variance $\sigma^2(0) = 2/s^2$ may be arbitrarily small for large values of $s$.

The mgf of the Laplace distribution is

$$M_0(\delta) = \exp(\delta m)/(1 - \left(\frac{\delta}{s}\right)^2).$$

Then, using equations (5.6) and (5.5b,c), one can show that

$$E^t[v] = m + 2t/(s^2 - t^2)$$

and

$$\sigma^2(t) = 1/(s-t)^2 + 1/(s+t)^2.$$

We can see that both current mean and variance tend to infinity as $t \to s$.

It is instructive to write the equations for the mean strategy $u(t) = E^t[v]$ of Malthusian model (5.3b) with different initial distributions (recall that according to equation (5.3a), $x(t) = x(0)\exp(u(t))$.)

a) Normal distribution:

$\frac{du}{dt} = \sigma^2, \sigma^2 = const$ is the variance of initial normal distribution; then

$u(t) = \sigma^2 t + u(0);$

b) Exponential distribution with the mean $1/s$:

$\frac{du}{dt} = \sigma^2 = \frac{1}{(s-t)^2}$ ; then

$$u(t) = 1 - \frac{1}{s} + \frac{1}{s-t}, u(0) = 1;$$

c) Truncated exponential distribution in the interval [0,1]:

$$\frac{du}{dt} = \sigma^2 = \left(\frac{1}{(s-t)^2} - \frac{e^{(s+t)}}{(e^s - e^t)^2}\right), u(0) = 0; \text{ then}$$

$$u(t) = 1 - \frac{1}{s} + \frac{1}{s-t} + \frac{1}{e^s - 1} + \frac{1}{e^{t-s} - 1};$$

d) Laplace distribution:

$$\frac{du}{dt} = \sigma^2 = \left(\frac{1}{(s-t)^2} + \frac{1}{(s+t)^2}\right), u(0) = 0; \text{ then}$$

$$u(t) = \frac{2t}{s^2 - t^2}.$$

One can see that equation (5.3b) has different forms and different solutions depending on the initial distribution of strategies and corresponding current variances.

Overall, as was shown above, the behavior of the strategy mean and variance (and all other statistical characteristics of the population) dramatically depends on the initial distribution of strategies in the population: even an arbitrarily small variance at the initial time moment cannot guarantee that it will not increase indefinitely, and even arbitrarily large initial variance may vanish over time. Therefore, in general one cannot assume that the variance of strategy distribution is constant; the current variance may change and be unknown, and how it will change over time is model and distribution dependent. Consequently, one cannot use the model of Darwinian dynamics without additional assumptions.

The method for overcoming the problem of unknown variance for a class of selection systems with the fitness that is linearly dependent on the strategies was described in (Kareva & Karev, 2019). Let us demonstrate the method on the example of system (4.3)-(4.4).

Instead of differential equation (4.3)-(4.4), we solve a formally infinitely-dimensional underlying system

$$\frac{dl(t,v)}{dt} = l(t,v)H(v,x) = l(t,v)[f(x) + vg(x)], \qquad (5.7)$$

$$x(t) = \int_V l(t,v)dv.$$

Assume that the initial values $l(0, u)$ are given and that $P(0, u)$ is the initial distribution of strategies. Let $M_0(\lambda) = \int_u \exp(\lambda u) P(0, u) du$ be the moment-generating function (mgf) of the initial distribution. Let us formally define the auxiliary variables $q$ and $p$ as solutions to the Cauchy problems

$$\frac{dp}{dt} = f(x), \, p(0) = 0, \quad \frac{dq}{dt} = g(x), \, q(0) = 0.$$

Then

$$l(t, u) = l(0, u) \exp(p(t) + uq(t))$$

and

$$x(t) = x(0) \exp(p(t)) M_0(q(t));$$

consequently, we obtain closed system for the auxiliary variables:

$$\frac{dp}{dt} = f(x(0) \exp(p(t)) M_0(q(t))), \, p(0) = 0,$$

$$\frac{dq}{dt} = g(x(0) \exp(p(t)) M_0(q(t))), \, q(0) = 0.$$

Solution to this system allows calculating all statistical characteristics of interest, including how the mean and variance of strategies in the population change over time.

The current distribution of strategies in the population is given by

$$P(t, v) = \frac{l(t,v)}{x(t)} = P(0, v) \exp(vq(t)) / M_0(q(t)), \tag{5.8a}$$

the current mgf is

$$M_t(\delta) = M_0(\delta + q(t))/M_0(q(t)), \tag{5.8b}$$

the mean strategies at $t$ moment are given by

$$E^t[v] = \frac{M_0'[q]}{M_0[q]}, \tag{5.8c}$$

and the variance at $t$ moment is

$$\sigma^2(t) = \frac{M_0''[q]}{M_0[q]} - \left(\frac{M_0'[q]}{M_0[q]}\right)^2, \tag{5.8d}$$

where the derivatives are taken with respect to $q$.

Using equations (5.8.a) or (5.8.b), it is easy to trace the evolution of initial distributions over time. In particular, if the initial distribution is normal with parameters $\mu, \sigma$, then at any time moment $t$ the distribution is again normal with parameters $\mu + 2q(t)\sigma^2, \sigma$. In this particular case, the variance of the current distribution is constant and does not change with time.

Now let us show that the normal distribution is the only case when the variance remains constant over time. Consider equation (5.8d) with the following initial conditions

$$M_0[0] = 1, M_0'[0] = E^0[v] = u(0).$$

It can be easily solved as

$$M_0(q) = exp\{\int_0^q \int_0^s \sigma^2(\tau) d\tau ds + u(0)q\}, \qquad (5.9)$$

which gives us a family of mgf's, depending on the behavior of the variance.

In particular, if one assumes that $\sigma^2$ is a constant, then

$$M_0(q) = exp(\frac{\sigma^2 q^2}{2} + u(0)q),$$

which is exactly the mgf of normal distribution.

Next, using the Price equation, it is easy to show that

$$\frac{d\sigma^2(t)}{dt} = m_3(t), \qquad (5.10)$$

where $m_3(t)$ is the 3rd central moment of the current distribution of strategies.

Indeed,

$$\frac{d\sigma^2(t)}{dt} = Cov[f(x) + vg(x), (v - E^t v)^2] =$$

$$= g(x) Cov\ [v, (v - E^t v)^2] = g(x)(E^t[v, (v - E^t v)^2] - (E^t[v]E^t[(v - E^t v)^2])$$

$$= g(x)(E^t[(v - E^t v + E^t v)(v - E^t v)^2] - E^t[v]E^t[(v - E^t v)^2] =$$

$$= E^t[(v - E^t v)^3] = m_3(t).$$

Hence, if $m_3(t) = 0$, e.g., if the distribution is symmetric, then $\frac{d\sigma^2(t)}{dt} = 0$; so, if $m_3(t) = 0$ for all $t$, then the variance is constant, and the distribution of $v$ is normal at any time.

To summarize, it was just proven that *if the current distribution of strategies for model (5.7) is symmetric at all times, then this distribution is normal, and its variance is a constant.*

## 6. Discussion

In their textbook (Vincent & Brown, 2005), the authors suggested an attractive approach for studying evolutionary dynamics of populations that are heterogeneous with respect to some strategy that affects the fitness of individuals in the population. The authors consider the approach as (more or less) universally applicable to models with any fitness function and any initial distribution of strategies, which is symmetric and has small variance. Here it was shown that the scope of the approach proposed by V&B is unfortunately much more limited.

The V&B approach is closely connected with the idea of "playing the field," developed by Maynard Smith (1982, p.23). According to this idea, the individual does not interact in a pair-wise fashion with other individuals; rather, the individual faces an opponent that is the population at large. As McGill and Brown (2007) wrote,

> "In this case we can think of the entire population playing the single strategy $U$. Under this interpretation we see $u$ as the strategy of a mutant individual or a focal individual, and $U$ is the resident strategy, or the strategy found among the $N$ individuals of the population (or $N-1$ with 1 for the mutant). Even if individuals in the resident population show variation, the playing-the-field approach can still work if the fitness of the target individual is well approximated by considering the average strategy of the resident population…. In the case of "playing the field", the payoff to an individual adopting a particular strategy depends not on the strategy adopted by one or a series of individual opponents but on some average property of the population as a whole, or some section of it."

Traditionally, an individual's fitness is defined as an average payoff under pairwise interaction of the individual with all other individuals in the population. In this case, the individual payoff essentially depends on the distribution of traits in the whole population. The individual payoff while "playing the field" depends only on the average trait, ignoring other statistical characteristics of the population (such as variance, etc.).

Here it was shown that the equation for the dynamics of the total population, as proposed by V&B or, more broadly, the playing-the-field approach, is applicable only if the fitness function is close to linear over the individual strategy. In general, fitness of an individual depends not only on the mean strategy of the population; rather, it can dramatically depend on the distribution of strategies in the population, even if the mean value of the strategy is the same at the initial time moment. This means that in the case of a general fitness function and trait distribution, the definition of individual fitness according to the "playing the field" concept is just incorrect.

The $2^{nd}$ equation of the V&B model for dynamics of the mean strategy assumes that the variance of strategy distribution is small and remains nearly constant over time. It was shown here that these conditions are in fact quite restrictive: depending on the initial distribution, variance can become arbitrarily large or tend to zero over time. In the simple case of linear fitness, the only case that guarantees a constant variance is when the initial distribution is normal. Notably, although the initial distribution in many cases can be approximated by normal distribution, in any realistic situation the range of strategies is bounded; then, the mean of strategies increases due to the Fisher' Fundamental theorem until the variance of the strategy distribution is non-zero. Therefore, the mean value tends to a maximal possible value of the strategy parameter, and so over time, the strategy distribution will tend to a distribution concentrated in this value. This of course means that the variance will be not constant but will tend to zero with time.

Overall, the model of Darwinian dynamics suggested and applied by V&B to a large number of different population models can unfortunately be used only subject to very limited conditions and can make incorrect predictions if these conditions do not hold.